\documentclass[twocolumn,floatfix, showpacs]{revtex4}

\usepackage[final]{epsfig}
\usepackage{amsmath}
\begin{document}
 
\title{Photon Emission from a Medium-Modified Shower Evolution}
 
\author{Thorsten Renk}
\email{thorsten.i.renk@jyu.fi}
\affiliation{Department of Physics, P.O. Box 35, FI-40014 University of Jyv\"askyl\"a, Finland}
\affiliation{Helsinki Institute of Physics, P.O. Box 64, FI-00014 University of Helsinki, Finland}


\begin{abstract}
Photons from the interaction of a highly energetic jet with a thermal medium are an important contribution to the total photon yield measured in ultrarelativistic heavy-ion collisions and also an important probe to study the medium degrees of freedom. Previously this contribution has often been computed in the context of a leading parton energy loss approximation. In this work, jet-medium interaction photons are instead estimated using  a medium-modified shower evolution model, where the energy degradation due to vacuum radiation prior to medium formation, the virtuality evolution of intermediate states and the photon emission from subleading shower partons is taken into account consistently. The results indicate that the leading parton energy loss approximation does not appear to work well for photon emission from jet-medium interaction.
\end{abstract}
 
\maketitle

\section{Introduction}

Photonic measurements constitute an important part of the set of observables considered to study the property of the Quantum Chromodynamics (QCD) medium created in ultrarelativistic heavy-ion collisions. As with dileptons, the other major class of electromagnetic (e.m.) observables, their value arises from  the relative smallness of the e.m. coupling $\alpha_{em}$ as compared to the strong coupling $\alpha_s$ which implies that any photon or lepton has a mean free path two orders of magnitude larger than the constituents of stronlgy interacting matter. Thus, once produced in a medium, e.m. probes escape practically without any re-interaction from the medium. 

Thermal photon emission as a tool to study the spacetime emission of heavy-ion collisions has been studied as early as in the SPS era \cite{ThomaRpt,Dinesh,Huovinen,MyPhotons} and has now reached a high degree of sophistication, involving state-of-the art modelling of the medium evolution in terms of event-by-event fluctuating or viscous hydrodynamics \cite{Dion,Rupa1,Rupa2} aiming to reproduce the measured angular momentum anisotropy coefficient $v_2$ \cite{PHENIX-v2}.

In a different momentum regime, hard direct photons produced from perturbative high $P_T$ partonic reactions in the medium are often considered as the 'golden channel' to measure parton-medium interactions, as measuring the photon allows to constrain the hard reaction kinematics very well \cite{gamma-h-XN,gamma-h-XN1,gamma-h-Renk,gamma-jet-Vitev,Bias}. 

In addition to thermal photons originating from the medium and hard photons produced in high $P_T$ reactions, there are additional sources of photons related to the interaction of hard partons with the medium, for instance the so-called jet conversion photon contribution \cite{ConversionPhotons} or the induced radiation contribution \cite{Zakharov}. Such photons are a relevant background to thermal photon observables, for instance they tend to dilute any photon $v_2$ contribution, but also dilute the clean kinematics of a $\gamma$-triggered hard correlation. However, since jet-thermal photons carry information about jet-medium interaction, they are interesting in their own right and for instance jet-tagging has been suggested to isolate their contribution \cite{JetTagging}.

In \cite{Turbide}, the various sources of photons have been computed and classified as follows: \emph{Prompt-direct photons} directly result from the hard process itself. In contrast, \emph{prompt fragmentation} photons are radiated in the final state parton shower of a quark or gluon  produced in a hard process. \emph{Jet-QGP} photons come from the interaction of hard partons with the medium such as the conversion reaction, while \emph{thermal} photons are produced by interactions among medium partons.

This classification is motivated by the energy loss picture of the interaction of hard partons with the medium, in which the passage of a high $P_T$ parton through the medium is treated as energy loss from the leading parton due to induced radiation, followed by vacuum fragmentation of parton shifted in energy outside the medium (see e.g. \cite{ASW,GLV}). However, a more modern understanding, driven by the need to understand fully reconstructed jet observables, is to see the energy loss picture an approximation to a medium-modified shower evolution in which vacuum and medium-induced radiation are not separable emission by emission but only on average (see e.g. \cite{Constraining} for a detailed discussion of the energy loss approximation). There are now several Monte-Carlo (MC) codes available to compute in-medium shower evolution \cite{JEWEL,YaJEM1,YaJEM2,Q-PYTHIA,MARTINI,AbhijitMC}.

Since photon emission through Quantum Electrodynamics (QED) processes is subject to the same radiation phase space considerations as the vacuum or medium-induced partonic QCD radiation, this changed view has pronounced implications also for jet-medium photons. In particular, in a medium-modified shower picture one can not separate prompt fragmentation from medium-induced radiation photons, and the uncertainty relation implies that a significant part of the QCD vacuum radiation occurs \emph{before} medium-induced photons can be generated, not after the hard parton exits the medium as assumed in the energy loss approximation. The aim of this work is to explore jet-medium photon emission in a medium-modified shower picture, here exemplified by the MC code YaJEM \cite{YaJEM1,YaJEM2}.

\section{Qualitative considerations}

As most in-medium shower codes, YaJEM is based on the vacuum shower code PYSHOW \cite{PYSHOW} which simulates the QCD evolution of a highly virtual initial parton as a series of $1\rightarrow 2$ splittings of a parent parton into two daughters with decreased virtuality. The evolution is carried out until a lower non-perturbative virtuality scale $O(1 \text{ GeV})$ is reached, at which point a non-perturbative hadronization model (such as the Lund model \cite{Lund}) is used. The vacuum fragmentation photon contribution can in this picture be obtained by allowing in addition to the QCD processes $q \rightarrow qg$ $g\rightarrow gg$ and $g\rightarrow q\overline{q}$ also the QED $1\rightarrow 2$ splitting $q \rightarrow q \gamma$, i.e. any parton which is charged can radiate electromagnetically if there is phase space available.

In a schematic way, jet-medium interaction photons can be classified as being produced in two different processes: \emph{Bremsstrahlung photons} are produced when interaction with the medium makes a radiation kinematically possible by creating radiation phase space (on-shell charges can not radiate real photons, but interaction with the medium can move the charge off-shell). Diagrammatically, the simplest such processes are $2\rightarrow 3 $ reactions in which a photon line is attached to a QCD $2\rightarrow 2$ scattering diagram. In contrast, \emph{conversion photons} are the results of the $2\rightarrow 2$ processes $qg \rightarrow q\gamma$ or $q\overline q \rightarrow g\gamma$ (with $q\overline q \rightarrow \gamma \gamma$ much suppressed due to the weakness of the e.m. coupling) in which partons change their identity rather than radiation phase space.

In previous works in the context of the energy loss approximation (e.g. \cite{ConversionPhotons,JetTagging,Turbide}), it has been assumed that bremsstrahlung and conversion photons can be computed, as the medium-induced radiation, for an on-shell quark with initially the full energy coming from the hard process, and that the fragmentation photon component can then be obtained by vacuum-fragmenting the quark after its energy has been shifted down due to its interactions with the medium.

However, quarks are created with a high initial virtuality. The Heisenberg uncertainty principle suggests that parametrically a virtual state lives for a time $\tau = E/Q^2$ where $E$ is the energy and $Q$ the virtuality of the quark. Taking, as assumed by vacuum parton showers, a lower perturbative scale of 1 GeV, this implies that even at RHIC 20 GeV quark jets have a parametric lifetime of $\sim 4$ fm during which the virtuality is above 1 GeV, i.e. leading shower partons may be quite far from being on-shell while they pass through the medium for a significant time. Since initial virtualities parametrically are $Q \sim E$, the first branchings take place at timescales $\ll 0.1$ fm, i.e. \emph{before} any medium can form. Thus, typically vacuum radiation degrades the energy of the leading quark even before it can interact with the medium, and the state encountering the medium, rather than being a single on-shell quark, is a shower of still highly virtual quarks and gluons. In particular, in such a shower there may be secondary charges, i.e. also a gluon jet gives in principle rise to photon production.

This has a weak impact on computations of leading hadron production, as what matters for the observation of the leading hadron is mainly the energy of the leading quark before hadronization, and the time ordering (i.e. whether vacuum radiation reduced the energy first and medium-induced radiation later or vice versa) is irrelevant. This is one of the reasons the energy loss approximation works so well for leading hadron production \cite{Constraining}. That the same approximation holds is less obvious for photon production --- it can be argued that for the balance of fragmentation photons and bremsstrahlung photons a similar argument holds, and that increased energy loss implies stronger QCD and stronger QED induced radiation, compensated by a fragmentation with reduced energy later. However it is not obvious that the momentum dependence of both contribution cancels. 

A strong impact is however expected for the conversion photon channel: Here, the time ordering, i.e. whether some energy has been lost before the medium is encountered or after is crucial --- the rate of conversion photons for given momentum is much reduced if the quark energy is degraded before the medium forms. Likewise, the conversion processes are enhanced by $t$ and $u$ channel singularities for near on-shell quarks \cite{ConversionPhotons} which are assumed to be screened by thermal masses $O(gT)$ (with $g$ the coupling constant and $T$ the medium temperature). If the singularities are screened by the much larger virtuality of intermediate shower states instead, the conversion cross section is much reduced.

In the following, we will investigate the strength of the individual contribution in the context of the in-medium shower code YaJEM.

\section{The model}

As described above, the MC code YaJEM is based on the PYSHOW algorithm \cite{PYSHOW} which in turn is part of PYTHIA \cite{PYTHIA}. It simulates the evolution from an  initial parton with virtuality $Q_i$  to a shower of partons at lower virtuality in the presence of a medium. In the absence of a medium, YaJEM by construction reproduces the results of PYSHOW. A detailed description of the model can be found in \cite{YaJEM1,YaJEM2,YaJEM-D}. Here the version YaJEM-DE is used \cite{YaJEM-DE} which is one of the best-tested theoretical models available for in-medium shower evolution and gives a fair account of a large number of high $P_T$ observables both at RHIC and LHC \cite{Constraining,A_J,RAA-LHC,A_J_edep}. 

The medium-modification of the shower evolution is implemented via the modification of radiation phase space. The medium itself appears only through transport coefficients $\hat{q}, \hat{e}$ which influence the kinematics (in terms of energy $E_a$ and virtuality $Q_a$ of intermediate virtual states $a$ as

\begin{equation}
\label{E-Qgain}
\Delta Q_a^2 = \int_{\tau_a^0}^{\tau_a^0 + \tau_a} d\zeta \hat{q}(\zeta)
\end{equation}

and 

\begin{equation}
\label{E-Drag}
\Delta E_a = \int_{\tau_a^0}^{\tau_a^0 + \tau_a} d\zeta \hat{e}(\zeta)
\end{equation}

where $\tau_a^0$ is the time at which a fluctuation $a$ is created $\tau_a$ is the lifetime of the virtual state as given by a randomized evaluation of the Heisenberg uncertainty relation.

Individual scatterings with the medium are not resolved in this framework. This makes the computation of the combined vacuum and bremsstrahlung photon contribution straightforward by permitting in addition to QCD also QED branchings during the shower evolution. However, the conversion photon contribution can not be computed without resolving the medium.

In order to get an estimate for the importance of the conversion photon contribution, we assume in the following that \emph{for the purpose of evaluating the conversion processes only}, the medium can be locally decomposed as a free gas of quarks and gluons at temperature $T$ (note that such an assumption does not lead to a good description of the measured nuclear suppression factor $R_{AA}$ \cite{JussiMCRP}, it is therefore at best an estimate).

In the following, we utilize the main result for the spectrum of conversion photons, given the distributions $f_q$ and $f_{\overline{q}}$ of hard quarks propagating through a medium with temperature $T$ given in \cite{ConversionPhotons} as

\begin{equation}
\begin{split}
\label{E-Conversion}
E_\gamma \frac{dN_\gamma}{d^3 p_\gamma d^4 x} = &\frac{\alpha \alpha_s}{4 \pi^2} \sum_{f=1}^{N_f} \left(\frac{e_{q_f}}{e} \right)^2\\
& \times \left[f_q (p_\gamma) + f_{\overline{q}}(p_\gamma) \right] T^2 \left[\ln \frac{4 E_\gamma T}{m^2} -1.916\right].
\end{split}
\end{equation}

Note that the momentum distribution of conversion photons is given by the distribution of converting quarks. Inserting the distribution $f_{q(\overline{q})}(p) = (2\pi)^3 \delta(x-x_0) \delta(y-y_0) \delta (z-ct) \delta^3(p)$, i.e. evaluating the expression for a single propagating virtual quark state at known position and momentum while setting the mass scale $m$ to the virtuality $m = Q$ allows to compute the the conversion probability of this state per unit time. This can be integrated from $\tau_a^0$ to $\tau_a^0+\tau_a$ to find the probability of the conversion during the lifetime of the virtual intermediate state, which can be sampled in YaJEM for each virtual quark state during the shower evolution. In the case of a conversion reaction, the propagating quark is changed into a photon and the shower evolution is carried out further for the remaining partons (i.e. in principle there can be more than one conversion in a shower, or there can be both conversion and fragmentation photons from the same shower, although in practice this is exceedingly rare).

\section{Results}

\subsection{Medium-modified fragmentation photons}

As found in \cite{YaJEM2}, the medium-modified fragmentation function computed in YaJEM obeys a scaling law that makes the shower largely independent of the precise functional form dependence of the transport coefficient $\hat{q}(\zeta)$ on the actual path $\zeta$ but only on the integral $\Delta Q^2 = \int d \zeta \hat{q}(\zeta)$ for most paths which can occur in a realistic medium evolution. This allows to characterize the strength of the medium modification by the single parameter $\Delta Q^2$.

\begin{figure}[!htb]
\begin{center}
\epsfig{file=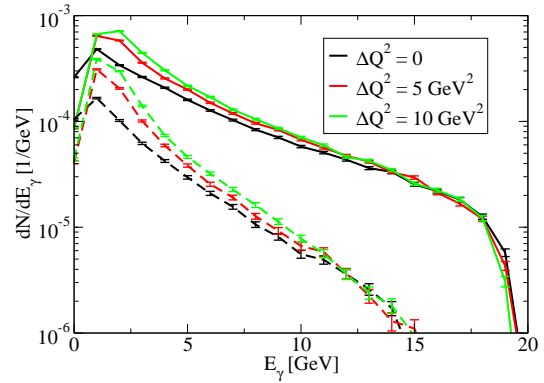, width=7cm}
\end{center}
\caption{\label{F-1} Medium-modified fragmentation photon yield from a 20 GeV parton evolved with YaJEM in vacuum and for two in-medium path with different strength of the total parton-medium interaction. Solid lines denote quark results, dashed lines denote gluon results.}
\end{figure}

Fig.~\ref{F-1} shows computed medium-modified fragmentation photon distribution originating from 20 GeV shower-initiating partons for two different strength of the medium evolution. Here, $\Delta Q^2 = 10$ GeV$^2$ approximates a long path through the center of a medium created at RHIC conditions, whereas $\Delta Q^2 = 5$ GeV$^2$ represents the average in-medium path. Overall, the effect of the medium is rather modest and  largely confined to low $E_\gamma$ which makes it difficult to observe (when convoluted with a pQCD parton spectrum as in the computation of a photon yield, dominantly the region above 15 GeV where differences vanish is probed). 

Fragmenting gluons give rise to a small photon component which comes from the splitting $g \rightarrow q \overline{q}$ early on in the shower evolution where one of the daughter quarks later undergoes an e.m. emission. The fact that a QED splitting is not possible from the shower initiator much depletes the distribution at high $E_{\gamma}$.

\begin{figure}[!htb]
\begin{center}
\epsfig{file=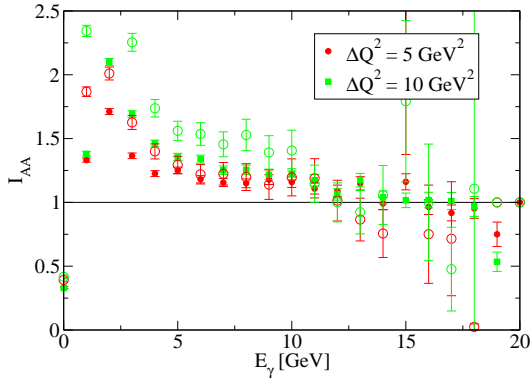, width=7cm}
\end{center}
\caption{\label{F-2} Ratio of medium-modified over vacuum fragmentation photon yield from a 20 GeV parton evolved with YaJEM  for two in-medium path with different strength of the total parton-medium interaction. Solid symbols denote quark results, open denote gluon results.}
\end{figure}

In order to bring out the differences between medium-modified and vacuum distribution better, we plot their ratio ($I_{AA}$) in Fig.~\ref{F-2}.

Again, the dominance of soft medium-induced e.m. radiation is clearly visible. Interestingly, the relative enhancement of the medium-modified yield is about a factor two higher for gluons, reflecting the fact that gluons always radiate photons by splitting into a $q\overline{q}$ pair. At high $P_T$ the relative enhancement is consistent with unity.

The medium-induced radiation spectrum depends on the energy scale of the shower initiating parton. Repeating the computation for fragmenting 200 GeV quarks and gluons yields no visible in-medium enhancement of the photon production within statistical errors.

\subsection{Conversion photons}

\begin{figure}[!htb]
\begin{center}
\epsfig{file=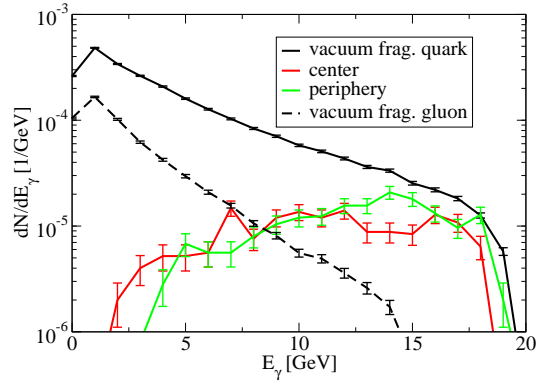, width=7cm}
\end{center}
\caption{\label{F-3} Yield of conversion photons resulting from a 20 GeV quark for two different paths through the medium, compared with the vacuum fragmentation photon yield.}
\end{figure}

Using the prescription outlined above, the estimate for the conversion photon yield from 20 GeV shower-initiating quarks is shown in Fig.~\ref{F-3} and compared with the vacuum fragmentation yield. 

Since there is no reason to assume that the conversion photon yield obeys a scaling law, the estimate is done for two different actual paths through a 2+1d ideal hydrodynamical simulation of RHIC central heavy-ion collisions --- one path from the medium center, and one from the periphery (i.e. a vertex displaced by 4 fm towards the surface).

The conversion photon yield is found to be broadly distributed in energy (note that the results of \cite{ConversionPhotons,JetTagging} or the direct evaluation of Eq.~(\ref{E-Conversion}) for the shower-initiating quark would expect it to be of the form $\sim \delta(E_{\gamma}-20\text{ GeV})$), indicating the strong role of vacuum radiation prior to the medium interaction. Furthermore, the yield is found to be smaller than the fragmentation yield throughout the whole energy range. While in \cite{ConversionPhotons} a strong conversion yield has to be obtained, this yield in the investigation here is  much reduced in the computation here due to the screening of the singularities by a large timelike virtuality $> 1$ GeV for virtual states as compared to screening by a thermal mass of order $\sim gT$. 

Somewhat surprisingly, the yield of conversion photons is relatively independent of the actual in-medium path. This is the result of an accidential  cancellation between the $T$ dependence of the conversion probability favouring a high-density medium and energy degradation by medium-induced radiation prior to conversion, favouring a lower density medium.

From Eq.~(\ref{E-Conversion}) one can expect that the shape of the photon spectrum reflects the shape of a quark distribution. This interpretation is blurred by the fact that the conversion photon yield is obtained as the result of a spacetime integration over an evolving medium, however the resulting shape is found to compare well with the quark distribution in the shower after an evolution time of about 2 fm. This timescale agrees reasonably well with the peak strength of medium-induced radiation (earlier times are screened by interference, later times are suppressed due to the rapidly decreasing medium density).

\section{Discussion}

The results obtained above suggest that the energy loss approximation can not be expected to give sufficiently precise results for photon emission from jet-medium interaction. In particular, the broad distribution of conversion photons seen in Fig.~\ref{F-3} given a quark with a well-defined initial energy suggests that jet-tagging of conversion photons as suggested in \cite{JetTagging} might not be feasible, and that the relative strength of the conversion photon contribution to the total should be re-evaluated in models beyond the energy loss approximation.

Fig.~\ref{F-1} suggests that overall the strength of the medium modification of photon emission is not large, and that the vacuum fragmentation contribution alone is a reasonably proxy for the total. A detailed investigation of the precise phenomenological consequences of this work is left for a future investigation. 

\begin{acknowledgments}
 
This work is supported by the Academy researcher program of the
Academy of Finland, Project No. 130472. 
 
\end{acknowledgments}

\end{document}